\documentclass[sigconf]{acmart}

\setcopyright{acmlicensed}
\copyrightyear{2025}
\acmYear{2025}
\acmDOI{XXXXXXX.XXXXXXX}
\acmConference[Conference acronym 'XX]{Make sure to enter the correct
  conference title from your rights confirmation email}{June 03--05,
  2025}{Woodstock, NY}
\acmISBN{978-1-4503-XXXX-X/2018/06}

\usepackage{xspace}
\usepackage{multirow}
\usepackage{booktabs}

\newcommand{\noind}[0]{\noindent}
\newcommand{\noindpar}[1]{\noind {\bf #1}}

\begin{document}
\abovedisplayskip=1pt
\abovedisplayshortskip=0pt
\belowdisplayskip=2pt
\belowdisplayshortskip=0pt
\title{AnomaSense: Anomaly-based Sensor Activation for Fine-Grained Human Activity Recognition}

\author{Xue Wang}
\orcid{0000-0002-4551-4932}
\affiliation{%
  \institution{University of California, Los Angeles}
  \city{Los Angeles}
  \state{CA}
  \country{USA}
}
\email{xw526@ucla.edu}

\author{Yang Zhang}
\orcid{0000-0003-2472-6968}
\affiliation{%
  \institution{University of California, Los Angeles}
  \city{Los Angeles}
  \state{CA}
  \country{USA}
}
\email{yangzhang@ucla.edu}

\renewcommand{\shortauthors}{Wang, et al.}
\def\systemname {\textit{AnomaSense}\xspace}

\begin{abstract}
Audio carries rich cues about human activities, and microphones are already built into most wearable devices. However, microphones also capture speech, and this privacy risk limits their use in Human Activity Recognition (HAR). We present \systemname{}, a sensor activation approach for wrist wearables that keeps the microphone off by default and turns it on for at most one second when an unsupervised anomaly detector flags an IMU segment that is likely to produce sound. The captured audio is further masked before it reaches the recognition model. We study 20 activities from 15 participants, organized into five groups in which activities share similar wrist motion but differ in the object or material involved. With IMU data alone, our recognition model reaches 78.98\% accuracy in leave-one-participant-out validation. With the short, masked audio windows added, accuracy reaches 96.89\% with no masking and stays above 86\% when 90\% of each one second audio window is removed. On the same data, the anomaly detector triggers the microphone with 86.46\% precision and 74.28\% recall relative to sound events. We also report a small preliminary check of automatic speech recognition on masked speech, which shows that contiguous masking degrades recognition far more than point-wise masking at the same masking ratio. Our evaluation is a controlled, offline feasibility study. We describe the threat model, what the approach does and does not protect, and the steps needed before deployment.
\end{abstract}



\begin{teaserfigure}
  \includegraphics[width=\textwidth]{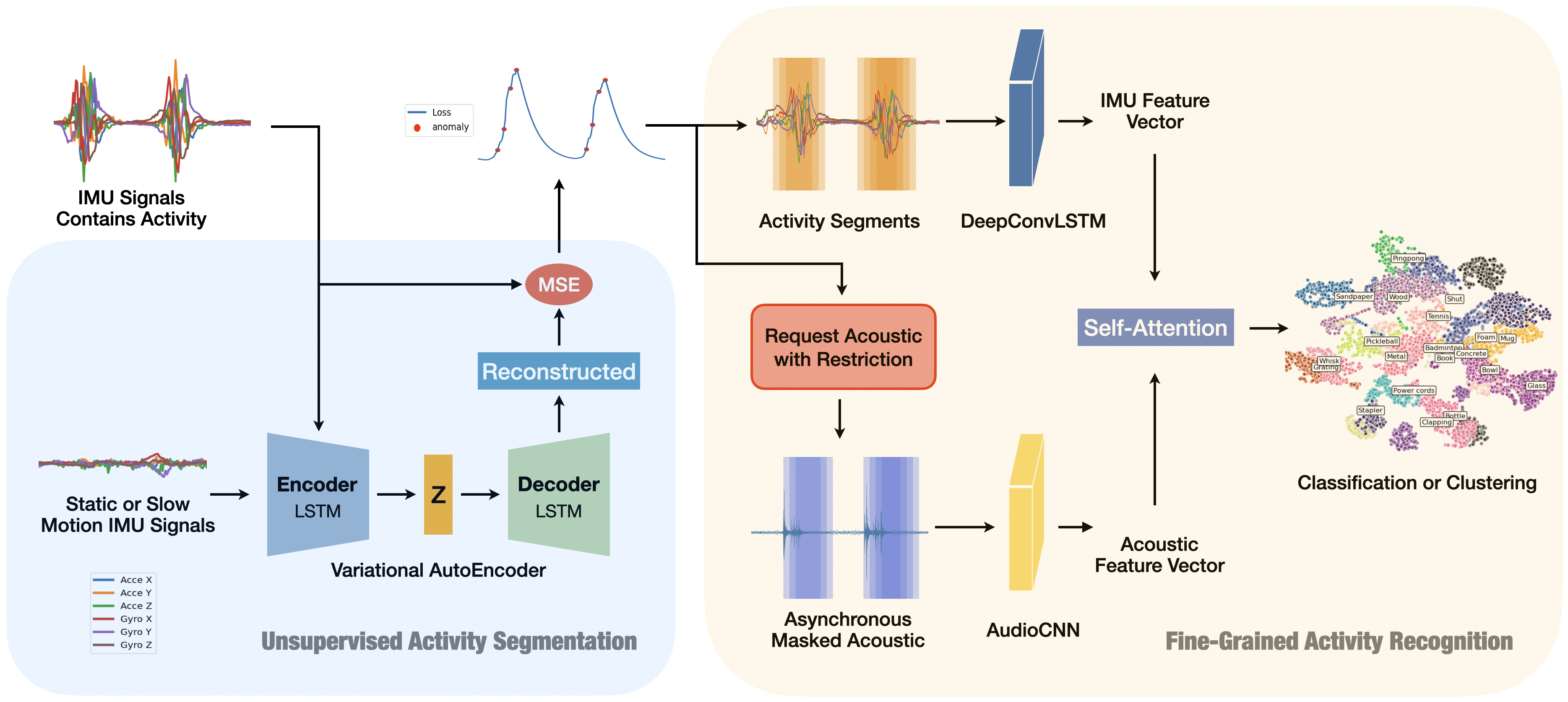}
  \caption{Overview of the \systemname{} pipeline. Left: unsupervised activity segmentation. An LSTM variational autoencoder trained only on IMU data from non-audible motion reconstructs the incoming six channel IMU stream. The reconstruction error is smoothed and thresholded, and points above the threshold are anomaly points. Right: fine-grained activity recognition. A one second IMU window centered on the anomaly point goes to a DeepConvLSTM branch, and a one second audio window requested at the anomaly point is masked and goes to an AudioCNN branch. The two feature vectors are fused with self-attention and passed to a classifier or, for unseen activities, a clustering step.}
  \label{fig:teaser}
  \Description{Block diagram with two shaded regions. The left region, labeled Unsupervised Activity Segmentation, shows six channel IMU signals entering an LSTM encoder, a latent vector Z, and an LSTM decoder. The reconstructed signal and the input are compared with mean squared error to give a loss curve with dots marking anomalies. The right region, labeled Fine-Grained Activity Recognition, shows highlighted IMU activity segments entering a DeepConvLSTM block, a box labeled Request Acoustic with Restriction leading to a masked audio waveform that enters an AudioCNN block, both feature vectors joined by a self-attention block, and a final scatter plot of clustered activity labels.}
\end{teaserfigure}

\maketitle

\section{Introduction}

Many human actions make sound. Footsteps, a mug set down on a table, a racket hitting a ball, a laptop lid closing. These sounds carry information that wrist-worn inertial sensors miss. An inertial measurement unit (IMU) records how the wrist moves, but two actions with nearly the same wrist motion can involve different objects and produce different sounds. Setting down a mug and setting down a bowl look alike to an IMU. So do a badminton swing and a tennis swing. Audio resolves these cases, and microphones are already present in most smartwatches.

The obstacle is privacy. A microphone that listens for activity sounds also captures speech \cite{malkin2019privacy, ur2020smartspeakers}. Prior work has responded in two ways. One line of work records audio and then removes or obfuscates speech content, for example by subsampling the audio to 1 kHz \cite{mollyn2022samosa}, filtering speech segments \cite{boovaraghavan2024kirigami}, or keeping only features \cite{laput2017synthetic}. The other line of work limits when the microphone is on. We follow the second line. Our goal is to keep the microphone off by default and turn it on for a short, bounded window only when the IMU suggests that a sound-producing action is happening.

In this paper we use the terms coarse and fine-grained in a specific way. A \textit{coarse} distinction is one that an IMU alone can make well, for example walking versus typing, or hammering versus placing a cup. A \textit{fine-grained} distinction is between activities that share nearly the same wrist motion and differ in the object, material, or surface involved, for example hammering wood versus hammering metal, or placing a glass versus placing a ceramic mug. Fine-grained distinctions are where audio adds the most and where IMU-only recognition struggles. Our activity set was built around fine-grained groups on purpose, and we return to what this design choice means for generality in Section \ref{sec: limitations}.

\systemname{} works as follows. An unsupervised anomaly detector, a variational autoencoder with LSTM encoder and decoder, is trained only on IMU data from non-audible everyday motion. At run time it reconstructs the incoming IMU stream and reports an anomaly point when the reconstruction error rises above a threshold. In Section \ref{sec: chronology} we argue that for many wrist actions the IMU signature of the action precedes or coincides with the sound it produces. An anomaly point therefore serves as a trigger. At each anomaly point the system requests one second of audio, masks it, and pairs it with a one second IMU window centered on the anomaly point. A multimodal recognition model then labels the activity. The detector does not need labels for the target activities, which lets the trigger generalize beyond the activity set used to train the recognizer. The IMU itself runs continuously, as it does in any smartwatch. The restrained sensor is the microphone.

We evaluate the approach with a controlled data collection from 15 participants performing 20 activities, arranged in five groups of four. Activities within a group share wrist motion and differ in the object or material. The evaluation is offline and simulates activation: both streams were recorded during the sessions, and the activation logic was applied afterwards so that we could measure how often the trigger fires at the right time and how the recognizer performs with only the audio it would have received. On this data the trigger reaches 86.46\% precision and 74.28\% recall against sound events identified from the audio energy. Recognition accuracy in leave-one-participant-out validation is 78.98\% with IMU only, 96.89\% with IMU plus unmasked one second audio windows, and 86.32\% when 90\% of each audio window is removed with contiguous masking. A preliminary check with an off-the-shelf speech recognizer, trained and tested on masked speech from four speakers, shows that contiguous masking degrades recognition much more than point-wise masking at the same ratio. We report this check because it informed our masking choice, not as a privacy guarantee.

We want to be direct about the scope of the claims. The activity set stresses the case where IMU is ambiguous and audio is not, so the gain from audio is larger than it would be on a mixed set of everyday activities. Sessions were controlled and quiet. The recognizer was trained and tested on our own data. The results show that anomaly-based microphone activation is feasible and that fine-grained recognition survives heavy masking of short audio windows. They do not show that the approach works in the wild. Section \ref{sec: threat model} states the threat model and deployment assumptions, and Section \ref{sec: limitations} lists what remains open.

Our contributions are:
\begin{itemize}
    \item A microphone activation approach that uses unsupervised IMU anomaly detection as the trigger, and an argument from the chronology of wrist motion and sound for why this trigger is reasonable.
    \item A recognition pipeline that pairs a one second IMU window with a one second, masked audio window, with results across masking types and ratios on 20 fine-grained activities from 15 participants.
    \item A first measurement of trigger accuracy, that is, precision and recall of microphone activation relative to sound events, which prior sensor activation work has not reported to our knowledge.
    \item A statement of the threat model for this style of sensor activation and an account of its limits, together with our code and dataset.
\end{itemize}

\section{Related Work}

\subsection{HAR with Wrist-Wearable Sensors}

\noindpar{Novel Sensors.} Human Activity Recognition (HAR) with wrist-wearable devices has been explored with many sensing modalities. Laser speckle imaging \cite{wang2024texturesight} detects grasped or touched objects to infer activities. Electromagnetic sensing, as in MagnifiSense \cite{wang2015magnifisense} and EM-Sense \cite{laput2015sense}, identifies electronic devices and touched objects. Maekawa et al. \cite{maekawa2012recognizing} used a finger-ring sensor to recognize handheld electrical devices such as cameras and phones. RFID sensors \cite{buettner2009recognizing} and RF-based systems such as Back-Guard \cite{yin2020back} have also been used for activity recognition. Wearable cameras are another option. First-person video supports recognition of activities of daily living \cite{pirsiavash2012detecting}, and wearable cameras carry their own privacy concerns for the wearer and for bystanders \cite{hoyle2014privacy}. A camera on the wrist would also need the same kind of activation control we study here for the microphone. In the rest of this section we focus on IMU and acoustic sensing, which are the two modalities our work uses.

\noindpar{IMU.} IMU signals have been used extensively for HAR. ViBand \cite{laput2016viband} samples a smartwatch accelerometer at high rate to sense bio-acoustic signals propagating through the wrist and classifies hand gestures such as flicks, claps, scratches, and taps. Laput and Harrison \cite{laput2019finegrained} extended this to 25 fine-grained hand activities using 4 kHz accelerometer data. iRoCo \cite{weigend2024iroco} estimates arm pose from IMU signals for human-robot interaction. CHARM-deep \cite{ashry2020charm} performs continuous, real-time activity recognition from smartwatch IMU data. ViObject \cite{chen2024viobject} uses smartwatch IMU data to identify untagged everyday objects from the vibrations caused by grabbing them.

\noindpar{Microphone.} Acoustic signals alone have also been used. Amento et al. \cite{amento2002sound} built a wrist-mounted bio-acoustic interface for fingertip gestures. Hambone \cite{deyle2007hambone} uses small piezoelectric sensors to detect sound for gesture-based interaction. AudioTouch \cite{kubo2019audiotouch} uses two piezoelectric elements as speaker and microphone to recognize hand gestures and touch force. The Sound of Touch \cite{mujibiya2013sound} uses transdermal low-frequency ultrasound for pressure-aware touch sensing on the body. Thomaz et al. \cite{thomaz2015inferring} ran an in-the-wild feasibility study to infer eating from ambient sound captured at the wrist.

\noindpar{IMU+Microphone.} Closest to our work are systems that combine IMU and acoustic signals. GestEar \cite{becker2019gestear} combines motion and audio on a smartwatch to recognize sound-emitting gestures such as snapping, knocking, and clapping. Bhattacharya et al. \cite{bhattacharya2022leveraging} used sound and wrist motion to detect daily living activities beyond hand gestures. Siddiqui and Chan \cite{siddiqui2020multimodal} used a 10-microphone array with accelerometer and gyroscope data to detect 15 hand gestures. These systems do not address the privacy cost of recording audio. AudioIMU \cite{liang2022audioimu} avoids audio at inference time by distilling an audio teacher into an IMU student, and reports a 4\% accuracy gain over IMU alone. That gain is modest, which suggests that audio information is hard to transfer fully into an IMU-only model and that some access to audio at inference time is still valuable. Our work asks how little audio access is enough.

\subsection{Privacy-Preserving Techniques for Audio-based HAR}
\noindpar{Signal-obfuscation approaches.} One family of methods records audio and then removes speech content. Larson et al. \cite{larson2011accurate} designed cough detection features that reconstruct cough sounds but not speech. PDVocal \cite{zhang2019pdvocal} tracks Parkinson's disease from non-speech body sounds only. Synthetic Sensors \cite{laput2017synthetic} extracts only the features needed for recognition from the audio stream. Liang and Thomaz \cite{liang2019audio} use embeddings learned from online video sound clips. VAX \cite{patidar2023vax} uses labels from off-the-shelf models to train privacy-sensitive sensors in situ. Kirigami \cite{boovaraghavan2024kirigami} filters out probable speech segments while maintaining HAR accuracy. PrivacyMic \cite{iravantchi2021privacymic} captures only inaudible frequencies on custom hardware. SAMoSA \cite{mollyn2022samosa} subsamples audio to 1 kHz and showed with a user study and public datasets that speech at this rate is largely unintelligible.

\noindpar{Sensor-activation approaches.} A second family limits when the microphone is on rather than what is done with the recording afterwards. The two families are complementary. Obfuscation reduces what a recording reveals. Activation reduces how much is recorded in the first place, so that less data exists to be processed, stored, or leaked. Two common activation patterns exist. In the first, a low-cost signal from the same sensor gates a higher-cost stage, as when a wake word such as ``OK Google'' or ``Hey Siri'' turns on full speech processing \cite{lopez2018alexa}. In the second, one sensor gates another. Infrared motion sensors that trigger wildlife cameras are a long-standing example \cite{swann2004infrared}. Apple Watch handwashing detection uses motion to start listening for the sound of running water \cite{apple2020handwashing}. SAMoSA \cite{mollyn2022samosa} uses a supervised IMU classifier to decide when to sample audio.

We propose two metrics for evaluating an activation approach, in addition to recognition accuracy. \textit{Sparingness} is how often and for how long the gated sensor is on. \textit{Trigger accuracy} is the precision and recall of activation relative to the events the gated sensor is meant to capture. To our knowledge, prior activation systems report the first, for example as the audio window length in SAMoSA, but not the second. We report both.

\noindpar{Relation to SAMoSA.} SAMoSA is the closest prior system and it is worth stating the differences plainly. SAMoSA triggers audio with a supervised IMU activity classifier, so the trigger is tied to the activities it was trained on. Our trigger is an unsupervised anomaly detector trained only on non-audible motion, so it does not need labels for target activities and, as Section \ref{sec: unseen} shows, fires on activities the recognizer has never seen. SAMoSA protects speech by subsampling to 1 kHz over a 3 second window and validated this with a formal study. We instead protect speech by limiting each audio request to at most one second and masking within that second, and we validate the masking with a small preliminary speech recognition check (Section \ref{sec: mask}) that is much weaker evidence than SAMoSA's study. Because we keep full-rate audio inside the short window, our pipeline can separate activities that differ mainly in high-frequency material sounds, which SAMoSA's coarse activity set does not target. Table \ref{tab: comparison} summarizes sensor usage and reported performance for the related systems and for our configurations. The numbers are taken from the respective papers and are not directly comparable, since the activity sets, participants, and protocols differ.

\begin{table*}[t]
\centering
\small
\resizebox{\textwidth}{!}{%
\begin{tabular}{lcccccc}
\toprule
\textbf{System} & \textbf{Audio rate} & \textbf{IMU rate} & \textbf{Sensor usage / label} & \textbf{Samples / label} & \textbf{Activities} & \textbf{Accuracy} \\
\midrule
AudioIMU \cite{liang2022audioimu} & no audio & 50 Hz & 10 s & 500 (IMU) & 23 coarse (e.g., write, chop) & 74.4\% \\
Fine-Grained Hand Activity \cite{laput2019finegrained} & no audio & 4,000 Hz & 3 s & 12,000 (IMU) & 25 fine-grained (e.g., clap, wash) & 95.2\% \\
GestEar \cite{becker2019gestear} & 11,025 Hz & 200 Hz & 300 ms & 3,308 (audio), 60 (IMU) & 9 fine-grained (e.g., knock left/right) & 97.2\% \\
SAMoSA \cite{mollyn2022samosa} & 1,000 Hz & 50 Hz & 3 s (audio), 2 s (IMU) & 3,000 (audio), 100 (IMU) & 26 coarse (e.g., drill, pour, knock) & 83.2\% (context independent) \\
\midrule
\systemname{}, no mask & 44,100 Hz & 50 Hz & 1 s (audio), 1 s (IMU) & 44,100 (audio), 50 (IMU) & \multirow{7}{*}{\shortstack{5 groups $\times$ 4 fine-grained\\ (e.g., hammering four materials,\\ swinging four rackets)}} & 96.89\% \\
\systemname{}, contiguous 50\%$\dagger$ & 44,100 Hz & 50 Hz & 500 ms (audio), 1 s (IMU) & 22,050 (audio), 50 (IMU) & & 92.33\% \\
\systemname{}, contiguous 80\%$\dagger$ & 44,100 Hz & 50 Hz & 200 ms (audio), 1 s (IMU) & 8,820 (audio), 50 (IMU) & & 88.19\% \\
\systemname{}, contiguous 90\%$\dagger$ & 44,100 Hz & 50 Hz & 100 ms (audio), 1 s (IMU) & 4,410 (audio), 50 (IMU) & & 86.32\% \\
\systemname{}, point-wise 50\%$*$ & 22,050 Hz effective & 50 Hz & 1 s (audio), 1 s (IMU) & 22,050 (audio), 50 (IMU) & & 95.55\% \\
\systemname{}, point-wise 80\%$*$ & 8,820 Hz effective & 50 Hz & 1 s (audio), 1 s (IMU) & 8,820 (audio), 50 (IMU) & & 95.17\% \\
\systemname{}, point-wise 90\%$*$ & 4,410 Hz effective & 50 Hz & 1 s (audio), 1 s (IMU) & 4,410 (audio), 50 (IMU) & & 94.27\% \\
\bottomrule
\end{tabular}%
}
\caption{Sensor usage and reported performance of related wearable IMU and acoustic HAR systems, and of \systemname{} configurations under different masking settings. Numbers for prior systems are as reported in the cited papers. Configurations marked with $\dagger$ use contiguous masking within the one second audio window, so the microphone is effectively on for the stated fraction of the second. Configurations marked with $*$ use point-wise masking, whose effect is comparable to sampling at the stated effective rate.}
\label{tab: comparison}
\end{table*}

\section{Principle of Operation}

\subsection{Scope, Threat Model, and Deployment Assumptions} \label{sec: threat model}

\noindpar{What we protect.} The asset is the content of speech, from the wearer and from bystanders, that a wrist-worn microphone would capture if it were on. The adversary is any party that can read what the microphone produces: the HAR application itself, a cloud service the application talks to, or an attacker who compromises either. We assume this adversary is adaptive. In particular, we assume the adversary knows the masking scheme and can train a speech recognizer on masked audio, which is why the check in Section \ref{sec: mask} trains the recognizer on masked data rather than on clean data.

\noindpar{How we protect it.} \systemname{} reduces the amount of audio that exists at all. The microphone is off by default. It is turned on only at anomaly points detected from IMU data, for at most one second per activation, and the second is further masked before it leaves the sensing layer. This is a data minimization measure. It bounds the exposure per activation and, because activations are tied to IMU anomalies rather than to time, it also breaks the continuity that speech recognition depends on. It does not make the retained audio content-free. A one second window can contain one or two words, and a masked window still leaks some acoustic information. Speech inpainting models can fill gaps in audio \cite{marafioti2019context}, and we expect that an adversary who collects many activations over a long period can recover more than our preliminary check suggests. We therefore describe the approach as reducing speech exposure, not eliminating it. The two masking techniques we study are compatible with, and should be combined with, downstream obfuscation such as subsampling \cite{mollyn2022samosa} or speech filtering \cite{boovaraghavan2024kirigami}.

\noindpar{What we do not protect.} IMU data is out of scope. The IMU runs continuously in our design, as it does in every smartwatch, and IMU data can itself support inferences about the wearer such as identity or gait \cite{kroger2019privacy}. Our design does not make that worse than a standard IMU-only HAR system, but it does not address it either. We also do not address inference from the activity labels themselves.

\noindpar{Where computation runs.} The anomaly detector has 0.539M parameters and needs 0.197 GFLOPs per one second window, evaluated every 0.1 seconds. This is within the range of models that run on current smartwatches, and it must run on the device for the design to make sense, since its job is to decide whether to turn the microphone on. The recognition model is larger and in our evaluation it ran offline on a laptop. If the recognizer runs off the device, the masked one second audio window leaves the device with each activation, and the protection is the bound on how much audio leaves, not where it is processed. If the recognizer runs on the device, only labels leave. Both are legitimate deployments of the same activation logic, and the privacy argument in this paper is about the first, weaker one.

\noindpar{A second benefit: power.} Keeping the microphone off and running the audio branch of the recognizer only at anomaly points also reduces energy use compared to continuous audio sensing. We did not measure power in this work and do not claim a number. We mention it because reviewers of an earlier version asked, and because it is the other reason a platform would want an activation signal of this kind.

\noindpar{What the evaluation is.} All results in this paper come from an offline simulation of activation. During data collection we recorded the IMU and the microphone continuously so that we could later measure how well the trigger aligns with sound events. We then applied the activation logic to the recorded IMU stream and gave the recognizer only the audio that the trigger would have requested. A live deployment would differ in latency, in the effect of ambient noise, and in the frequency of false triggers during unscripted daily activity. We discuss each of these in Sections \ref{sec: discussion} and \ref{sec: limitations}.

\subsection{Chronological Patterns of Acoustic and IMU Signals} \label{sec: chronology}

The activation idea rests on a simple observation. For a wrist action that makes a sound, the motion that causes the sound is recorded by the IMU before or at the moment the sound is produced. Closing a laptop lid is a motion that ends in a thud. Placing a cup on a table is a descent that ends when the table stops the cup. A racket swing accelerates until the racket meets the ball. In each case the IMU signature of the action, and in particular the sharp change in velocity, is a usable cue for when the sound will occur. An ideal activation approach would turn the microphone on at that moment and off shortly after.

The relationship is not the same for every action. For impulsive events such as clapping or hammering, the sound is short and arrives at the moment of the velocity change. For actions with a rebound or a follow-through, such as a racket swing, the sound arrives near peak velocity and the IMU signature continues afterwards. For repetitive actions such as whisking eggs, each wrist rotation produces a small sound and the IMU shows a series of micro-actions, so any anomaly point in the series is a reasonable place to listen. For actions with continuous friction sounds such as scrubbing with sandpaper, the sound and the IMU signal overlap for the whole action, and the timing of the trigger matters less as long as it falls inside the action.

There is also a class of activities for which this approach is a poor fit. Electrical appliances such as sanders or blenders emit sound continuously and independently of the wrist motion, so there is no moment that the IMU can predict. We excluded such activities from our study, since including them would reward the audio branch without testing the activation idea. This exclusion is a scope decision and it limits generality, which we return to in Section \ref{sec: limitations}.

A practical consequence of the chronology is the question of latency. If the trigger fires at the anomaly point and the microphone opens at that moment, sound that arrives before the anomaly point is missed. For impulsive sounds this can mean missing the onset. Our detector evaluates a one second window every 0.1 seconds, and the anomaly point is placed at the smoothed error peak inside that window. In our data the IMU error typically starts rising before the sound arrives, as Figure \ref{fig: activity segmentation illustration} shows for most activities, so a one second audio window starting at the anomaly point usually contains the sound. The recall of 74.28\% reported in Section \ref{sec: segmentation results} is partly a measure of how often this timing fails. We did not measure end-to-end latency on the watch, and a live system would need to add the detector's inference time to this picture.

\subsection{Preliminary Check: Masking and Automatic Speech Recognition} \label{sec: mask}

Within each one second activation we further mask the audio. We considered two simple masking methods. Both are defined by a mask factor between 0 and 1 that gives the fraction of samples removed.

\subsubsection{Point-wise masking}
\begin{figure}[t]
    \centering
    \makebox[0pt]{\includegraphics[width=\columnwidth]{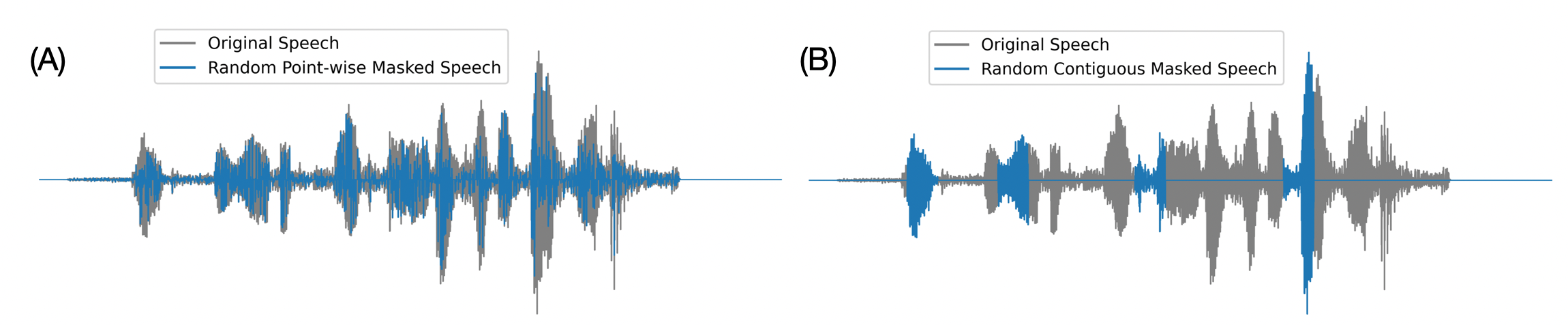}}
    \vspace{-0.12cm}
    \caption{Two masking methods applied to a speech waveform. (A) Point-wise masking sets individual samples to zero at random. (B) Contiguous masking sets one continuous segment of the window to zero, starting at a random position.}
    \Description{Two panels of audio waveforms. Panel A shows a speech waveform where scattered individual samples have been zeroed, leaving a waveform with the same overall shape but visible gaps at the sample level. Panel B shows the same waveform with one long continuous block set to zero, leaving speech only outside that block.}
    \label{fig: masking} 
\end{figure}

Each sample in the one second window is set to zero independently with probability equal to the mask factor, a Bernoulli process. This is memoryless, and its effect on the signal is similar to random subsampling. A 90\% point-wise mask keeps on average one sample in ten, which is comparable to sampling at 4,410 Hz. Figure \ref{fig: masking} (A) shows an example.

\subsubsection{Contiguous masking}
A start position is drawn uniformly at random within the window, and one contiguous segment of length equal to the mask factor times the window length is set to zero, wrapping at the window boundary. With a 90\% contiguous mask, the microphone is effectively on for 100 ms of the second. Figure \ref{fig: masking} (B) shows an example. In a deployment, contiguous masking corresponds to shortening the activation rather than to post hoc processing, which is why we treat it as the more interesting of the two.

\subsubsection{Recognition study of masked speech}
To choose between the two methods we ran a small check. Four participants, two of whom were women, each read 150 sentences of eight distinct words. Each participant read 50 sentences in each of three sensor placements: an iPhone 12 mini strapped to the right wrist with Velcro to stand in for a smartwatch, the same phone mounted on a helmet to stand in for a headset, and the phone lying on a table at random angles and distances to stand in for a smart speaker. We used the phone rather than the Pixel Watch used elsewhere in the paper because the watch app we built streams audio for recording but was not set up for this reading task. The check is about speech content, and we do not think the device changes the conclusion, but the inconsistency is a weakness and we list it in Section \ref{sec: limitations}.

We fine-tuned the audio model from Auto-AVSR \cite{ma2023auto}, which performs well on the LRS2 and LRS3 benchmarks, for 60 epochs with a learning rate of 5e-5 and a CTC loss weight of 0.1. Training and test sets were masked with the same method and factor, so the recognizer had every opportunity to adapt to the mask. We ran leave-one-speaker-out validation across the four participants. Table \ref{tab: mask asr} reports character error rate (CER) and word error rate (WER). The factors tested differ between the two methods because we swept each method over the range where its error started to move. Point-wise masking changed little below 80\%, and contiguous masking had already pushed WER near 70\% at 50\%.

\begin{table}[b]
\centering
\begin{tabular}[t]{cccc}
\toprule
\textbf{Method}&\textbf{Masked Data}&\textbf{CER}&\textbf{WER}\\
\hline
No Mask & - & 11.54\% & 15.46\% \\
\hline
\multirow{4}{*}{Point-wise Mask} & 80\% & 16.57\% & 22.70\% \\
                                 & 90\% & 30.64\% & 39.00\% \\
                                 & 95\% & 31.53\% & 40.33\% \\
                                 & 97\% & 39.11\% & 48.50\% \\
\hline
\multirow{3}{*}{Contiguous Mask} & 50\% & 56.62\% & 69.41\% \\
                                 & 80\% & 64.65\% & 79.61\% \\
                                 & 90\% & 64.93\% & 78.64\% \\
\bottomrule
\end{tabular}
\caption{Speech recognition error under different masking methods. Recognizer trained and tested on masked speech, four speakers, leave-one-speaker-out.}
\label{tab: mask asr}
\end{table}

Contiguous masking hurt recognition far more than point-wise masking. Removing 50\% of samples contiguously raised CER from 11.54\% to 56.62\% and WER from 15.46\% to 69.41\%. Removing 97\% of samples point-wise raised WER only to 48.50\%. When the authors listened informally to the masked files, the point-wise masked audio sounded noisier to the ear, yet the recognizer handled it well. We did not run a listening study with participants, and we do not report human intelligibility.

We draw two conclusions and note their limits. First, a recognizer trained on the mask recovers a lot from random sample dropout, so subsampling on its own is a weaker protection against an adaptive adversary than it sounds. Second, removing contiguous time, which is what a shorter microphone activation does, damages speech recognition more per sample removed. This is the reason our activation approach limits duration and frequency of microphone use rather than sampling rate. The check has four speakers, one recognizer, read speech, and quiet rooms. It is enough to motivate the design choice and not enough to characterize privacy, and we do not use it for any stronger claim.

\section{Implementation}
\subsection{Activity Segmentation}
Conventional HAR pipelines slide a fixed window over the sensor stream and classify each window \cite{liang2022audioimu, bhattacharya2022leveraging, valarezo2017human, konak2020imu}. The window length is a trade-off. A short window may not contain enough of the action, and a long window may contain more than one action \cite{ashry2020charm}. For our purpose the window also has a cost that a fixed slider does not account for: every window with audio is a microphone activation. We therefore segment the IMU stream first and request audio only for segments the detector marks as anomalous.

We use unsupervised anomaly detection on the multivariate IMU time series \cite{xu2021anomaly}. Among the available families, density estimation \cite{breunig2000lof}, clustering \cite{scholkopf2001estimating}, prediction \cite{pena2013anomaly, hundman2018detecting}, and reconstruction \cite{ringberg2007sensitivity, hsieh2019unsupervised, park2018multimodal, li2019mad, geiger2020tadgan}, we chose a reconstruction-based model. Reconstruction error is a reasonable score for contextual and collective anomalies \cite{wong2022aer}, which is what an activity embedded in a stream of ordinary motion is. The model is trained only on IMU data from motion that does not produce sound. At run time, motion that the model has not learned to reconstruct produces a high error, and we treat a high error as the trigger.

A reviewer of an earlier version asked whether modeling only ordinary motion is enough, since sound-producing actions share features with ordinary ones such as arm extension or body rotation. It is not enough to guarantee a clean trigger, and we do not claim that. The detector fires on any motion it reconstructs poorly, including some that produce no sound. The precision of 86.46\% in Section \ref{sec: segmentation results} is the measurement of how often that happens on our data. What the design buys is that the trigger does not depend on labels for the target activities, which is what lets it fire for activities the recognizer has not seen (Section \ref{sec: unseen}).

\subsubsection{Data collection of ``normal'' IMU signals}\label{sec:normaldatacollection}
We recruited 10 participants, none of whom took part in the later recognition study, to record ordinary non-audible motion with a Google Pixel Watch on the dominant wrist. We chose 10 activities that involve the arm and wrist, occur often in daily life, and make no sound: waving the arm, shaking hands, pressing a button on the watch touchscreen, stretching the arms, rotating the body, blotting the face with a hand, rubbing the eyes, putting on glasses, tying shoelaces, and putting a hand in a pocket. The list is not exhaustive. It was meant to cover common arm motions of varied amplitude so that the detector does not fire simply on large movements. Each participant repeated each activity five times and returned the arm to the side of the body between repetitions. We recorded linear accelerometer and gyroscope data at 50 Hz, which is the rate used by the related systems in Table \ref{tab: comparison} and is standard for wrist HAR. In total we collected 32.09 minutes of data.

\subsubsection{IMU signal anomaly detection} \label{sec: anomaly detection}
The left half of Figure \ref{fig:teaser} shows the detector. The input has six channels, three from the accelerometer and three from the gyroscope, and no audio. We use an LSTM variational autoencoder (LSTM-VAE), a standard architecture for multivariate time series anomaly detection \cite{chalapathy2019deep, valarezo2017human}. The encoder is a single-layer unidirectional LSTM that maps a window of IMU samples to the mean and log variance of a Gaussian latent vector. The decoder is a single-layer unidirectional LSTM that maps a sample from the latent distribution back to a window of six channel IMU values. We did not modify the standard LSTM cells. The loss is the negative evidence lower bound, computed as the mean squared reconstruction error plus the Kullback-Leibler divergence between the latent distribution and a unit Gaussian. The detector has 0.539M parameters and needs 0.197 GFLOPs per window. Model hyperparameters such as hidden size and latent size are given in the released code.

At inference we compute the mean squared error between the input window and its reconstruction and smooth the error with an exponentially weighted moving average over 0.1 seconds. We evaluate a one second window every 0.1 seconds. A step of 0.1 seconds was chosen because natural wrist actions do not produce more than one sound event within 0.1 seconds. Within each window we apply an adaptive threshold, and we also require the smoothed error to exceed an absolute floor of 0.5, chosen from the distribution of errors on static and slow motion in the training data, so that small fluctuations do not trigger the microphone. A time step at which the smoothed error exceeds both thresholds is an anomaly point.

\subsubsection{Assembly of input data for the recognizer}
At each anomaly point the system requests one second of audio starting at that point, and pairs it with the IMU samples from 0.5 seconds before to 0.5 seconds after the point. The two windows are therefore offset by half a second, which follows from the chronology in Section \ref{sec: chronology}: the IMU cue is centered on the action and the sound follows. The audio window is then masked with one of the two methods from Section \ref{sec: mask} before it enters the recognizer. With contiguous masking the effective microphone-on time per activation is the unmasked fraction of the second, from 500 ms at 50\% down to 100 ms at 90\%. We chose one second windows because the impulsive sounds in our activity set last well under a second and the window gives some margin for timing error. We did not run a full window length ablation. Section \ref{sec: window} reports what we observed with a 0.5 second window.

\subsection{Activity Recognition}
The right half of Figure \ref{fig:teaser} shows the recognizer. It has three input branches, for accelerometer, gyroscope, and audio, and a fusion stage. For the IMU branches we use DeepConvLSTM \cite{ordonez2016deep} without architectural changes. Each branch has four 2D convolutional layers, each followed by batch normalization, and the outputs of the accelerometer and gyroscope branches are concatenated and passed through a fully connected layer and then an LSTM layer. For the audio branch we compute a log-mel spectrogram from the masked one second window using a 1024 sample analysis window, a hop of 320 samples, and 64 mel bins. Masked samples are zeros and are treated as such by the spectrogram. The spectrogram goes to an AudioCNN following the design used in AudioIMU \cite{liang2022audioimu}.

The three branch outputs are concatenated and passed through a single self-attention layer, so that the model can weight the modalities per input, and then through a fully connected classifier. The full recognizer has 0.369M parameters and needs 97.488M FLOPs per input, which is small for a multimodal model. We train with Adam and cross-entropy loss. 

\section{Data Collection} \label{sec: data}
We collected a dataset from 15 participants to evaluate both the trigger and the recognizer. Each participant wore a Google Pixel Watch running a custom app that streams synchronized IMU and microphone data over the network to a MacBook Pro (Apple M2 Pro, 2023), where both streams were saved to files. The microphone was recorded continuously during the sessions. This is the opposite of how the system would run, and it was done on purpose. To measure whether the trigger fires at the right time we need to know when sounds occurred, including sounds the trigger missed, and that requires a complete audio record. Activation was simulated afterwards on the recorded data as described in Section \ref{sec: threat model}.

\subsection{Activity Set Design} \label{sec: activity design}
We designed 20 activities in five groups of four. The design principle is that activities within a group should share wrist motion and differ in the object, material, or surface, so that an IMU-only recognizer confuses them and audio is needed to separate them. Groups differ from each other in motion, so across-group distinctions are coarse in the sense defined in the introduction and are mostly solved by the IMU. This is a stress test for the fine-grained case, not a sample of daily life, and the accuracy numbers should be read that way.

The \textit{sports} group uses four rackets or paddles to hit their respective balls: badminton, pickleball, ping-pong, and tennis. The four swings are similar at the wrist and the impact sounds differ in pitch and duration. The \textit{materials} group hammers four surfaces: foam, metal, wood, and concrete brick. This group is the most artificial of the five. We included it because it isolates the material sound while holding the motion almost fixed, and because the surface being struck is a plausible thing to want to know in a workshop or construction context. The \textit{dining utensils} group places four objects on a table: a plastic bowl, a glass goblet, a water bottle, and a ceramic mug. Lifting and placing are the same motion and the contact sounds differ. The \textit{work and study} group has four office actions: opening and closing a book, opening and closing a laptop, plugging and unplugging a power cord, and stapling paper. These are less motion-matched than the other groups and were included to have some activities whose sound is intermittent, for example a laptop that makes a sound on closing but not on opening. The \textit{daily} group has four household actions: whisking eggs in a bowl, grating carrots, scrubbing wood with sandpaper, and clapping. These cover the repetitive and continuous friction sound patterns described in Section \ref{sec: chronology}.

The set is not balanced across genders, ages, or occupations, and the sports group in particular assumes familiarity with racket sports. We discuss this in Section \ref{sec: limitations}.

\subsection{Apparatus and Procedure}
The app records the linear accelerometer and gyroscope at 50 Hz and the microphone at 44,100 Hz. We kept audio at the device's full rate so that the masking analysis could control the effective audio resolution in software, and so that material sounds with high frequency content were available to the audio branch. The 50 Hz IMU rate matches the related systems in Table \ref{tab: comparison}.

Each participant wore the watch on the dominant wrist and was asked to perform each activity naturally. For the sports activities, participants who did not play the sport were asked to swing at the ball as they would in play, as many times as needed to get five hits. Participant 13 could not complete the badminton activity because they could not make contact with the shuttlecock, so that participant has no badminton data.

A session lasted about 30 minutes and had two parts. In each part the participant performed all 20 activities, each five times in a row. Between the two parts the participant removed the watch and put it back on, so that the two parts have slightly different watch placement. Each activity therefore has ten repetitions per participant, apart from the missing badminton data for one participant. Because activities were performed in dedicated blocks, the sessions contain little of the unscripted ordinary motion that the anomaly detector is meant to ignore, and the evaluation of false triggers in Section \ref{sec: segmentation results} is limited to the motion that occurs within and between repetitions. Section \ref{sec: limitations} discusses this.

\section{Evaluation}
We evaluate the trigger and the recognizer separately, both on the dataset of Section \ref{sec: data}.

\subsection{Trigger Accuracy: Activity Segmentation Results} \label{sec: segmentation results}
\noindpar{Ground truth.} To score the trigger we need to know when sounds occurred. We derived sound events from the continuously recorded audio rather than from video or manual labels. Within each one second buffer of audio we computed the mean and standard deviation of the signal and marked any sample above the mean plus two standard deviations as part of a sound event. This gives an energy-based event timeline that is independent of the IMU. A detected anomaly window that overlaps a sound event is a true positive. A sound event with no overlapping anomaly window is a false negative. An anomaly window with no sound event is a false positive. This last case includes exactly the situation a reviewer of an earlier version raised, a racket swing that misses the ball: the IMU shows an anomaly, no sound follows, and our metric counts it against the trigger. Because the ground truth comes from the audio energy, quiet sound events such as a book opening gently can be missed by the threshold, which would make some correct triggers look like false positives. We did not hand-verify the event timeline, and this is a limitation of the measurement.

\noindpar{Result.} Over all participants and activities the trigger reached a precision of 86.46\% and a recall of 74.28\%. Read against the threat model, precision is the fraction of microphone activations that captured a sound event, and one activation in seven captured nothing useful. Recall is the fraction of sound events for which the microphone was on, and one event in four was missed. Misses come from two sources that we could not separate cleanly: the detector not firing, and the detector firing late so that the one second audio window started after the sound. The sports group, where the anomaly point tends to sit in a long swing, contributes many of the misses.

\begin{figure}[t]
    \centering
    \makebox[0pt]{\includegraphics[width=\linewidth]{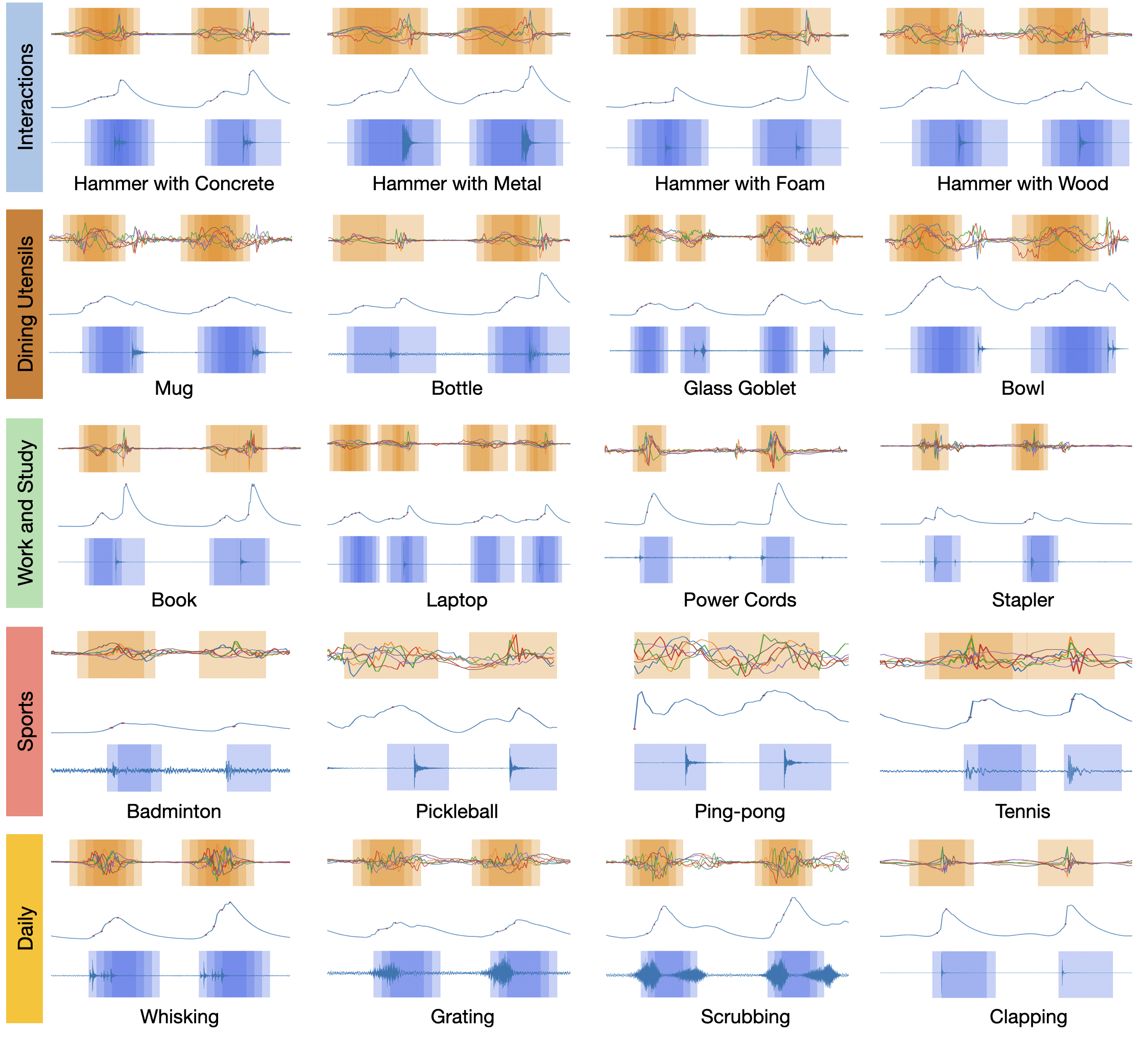}}
    \caption{Activity segmentation output for the 20 activities, one example recording each, grouped by row. For each activity, the top trace shows the six IMU channels with the one second IMU windows selected by the detector shaded in orange. The middle trace is the smoothed reconstruction error, with small dots marking detected anomaly points. The bottom trace is the audio waveform with the requested one second audio windows shaded in blue. The dots are small at this scale. For clearer illustration this figure uses 0.5 second windows.}
    \Description{A five by four grid of small multi-trace plots. Rows are labeled Interactions, Dining Utensils, Work and Study, Sports, and Daily. Each cell has three stacked traces: colored IMU lines with orange shaded regions, a single blue loss curve with peaks, and an audio waveform with blue shaded regions. In most cells the shaded regions in the top and bottom traces line up with the peaks of the loss curve and with the bursts in the waveform. Cells for the sports row show longer, less peaked loss curves and shaded regions that cover broader parts of the traces.}
    \label{fig: activity segmentation illustration} 
\end{figure}

Figure \ref{fig: activity segmentation illustration} shows one recording per activity. For impulsive activities such as hammering, stapling, and clapping, the error peaks are sharp and the audio windows sit on the sound bursts. For the dining utensils the peaks are lower and broader because placing an object is a slower motion, and the audio window still tends to contain the contact sound. For the sports group the error stays elevated through the swing, the anomaly points are less well defined, and the audio windows are wider relative to the sound. This matches the chronology discussion in Section \ref{sec: chronology} and is where the timing argument is weakest.

\begin{figure}[b]
    \centering
    \makebox[0pt]{\includegraphics[width=\columnwidth]{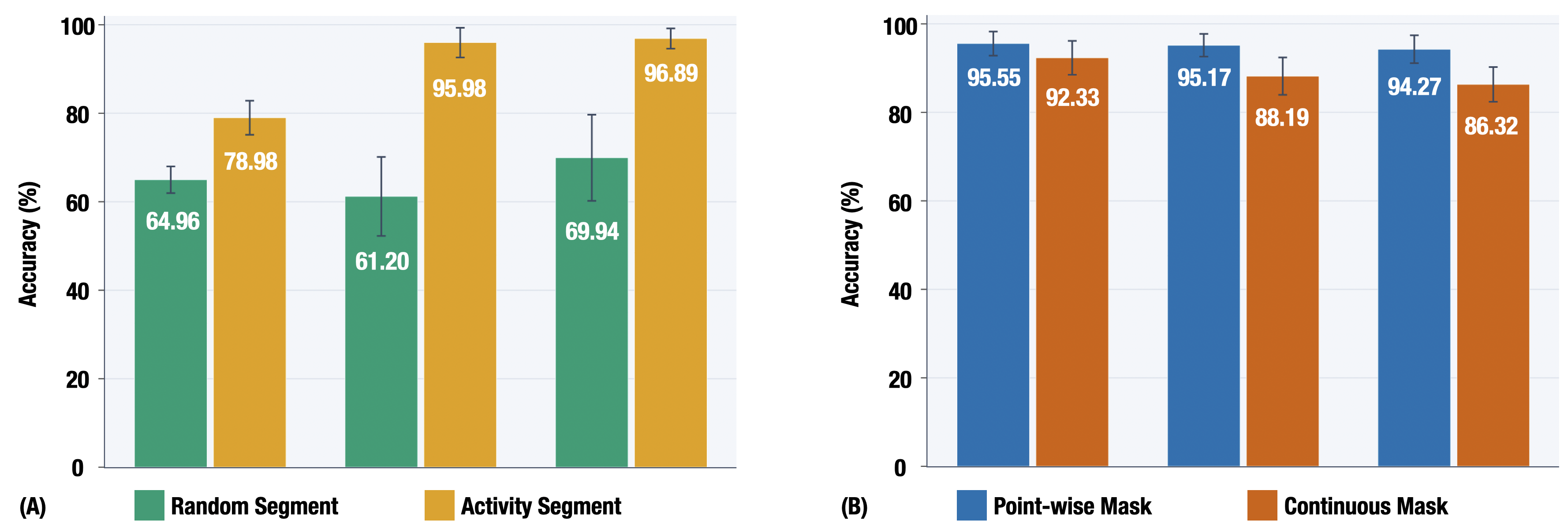}}
    \caption{Leave-one-participant-out recognition accuracy. (A) Trigger-selected activity segments versus randomly placed segments, for IMU only, audio only, and IMU plus audio, without masking. (B) Point-wise versus contiguous masking at 50\%, 80\%, and 90\% mask factors, IMU plus audio. Error bars are standard deviation across held-out participants.}
    \Description{Two grouped bar charts. Chart A has three groups labeled IMU Only, Audio Only, and IMU plus Audio, each with a green bar for random segments and a yellow bar for activity segments. The yellow bars are higher in every group, with values 78.98, 95.98, and 96.89 versus 64.85, 61.19, and 69.28. Chart B has three groups labeled 50 percent, 80 percent, and 90 percent mask, each with a purple bar for point-wise masking and a red bar for contiguous masking. Purple bars read 95.54, 95.16, and 94.28. Red bars read 92.33, 88.19, and 86.32, decreasing as the mask factor increases.}
    \label{fig: results} 
\end{figure}

\noindpar{Does segmentation help recognition?} To check that the trigger picks informative windows, we compared recognition accuracy on trigger-selected windows against accuracy on windows placed at random within the same recordings, matched in number per activity and participant, using leave-one-participant-out validation. Figure \ref{fig: results} (A) shows the result. Trigger-selected windows gave higher accuracy for every modality: 14.02 points higher with IMU only, 34.78 points higher with audio only, and 26.95 points higher with IMU and audio. We want to be clear about what this comparison shows. Random placement is a lower bound, not a competitive baseline. For impulsive sounds it is expected that a random one second window often misses the sound entirely, so the audio-only gap in particular is large by construction. The comparison confirms that the trigger lands on the sound often enough to matter for recognition. It does not show that our detector is better than a fixed sliding window with overlap, or than a simpler energy-based motion trigger, and we did not run those comparisons.

\subsection{Human Activity Recognition Results} \label{sec: HAR results}

\noindpar{Protocol.} We used leave-one-participant-out validation. For each fold, one participant's data was the test set, two other participants chosen at random formed a validation set for model selection, and the remaining 12 participants formed the training set. Because the number of anomaly points differs across activities and participants, we compute accuracy per activity and report the mean over the 20 activities, so that frequently triggered activities do not dominate the number. All differences reported below are differences in this mean accuracy across folds. We did not run statistical tests, and when we say one setting did better than another we mean the mean was higher, with the standard deviation across folds given in Table \ref{tab: mask har} for reference.

\begin{figure}[b]
    \centering
    \makebox[0pt]{\includegraphics[width=\columnwidth]{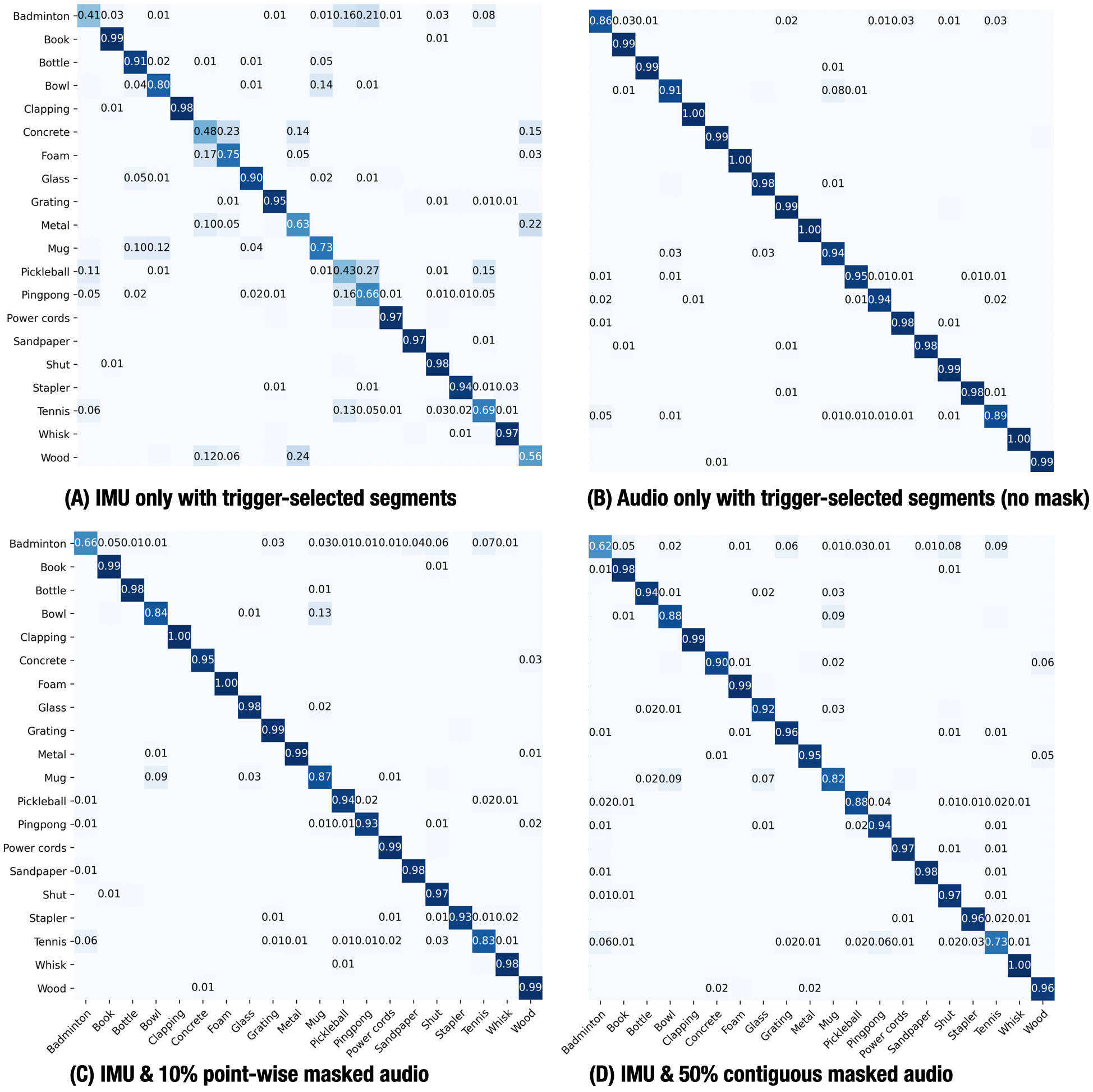}}
    \caption{Confusion matrices for four settings, rows are true labels and columns are predictions, 20 activities grouped as in Section \ref{sec: activity design}: (A) IMU only with trigger-selected segments; (B) audio only, no mask, trigger-selected segments; (C) IMU plus audio with a 10\% point-wise mask; (D) IMU plus audio with a 50\% contiguous mask. The label \textit{shut} denotes the sound produced when closing the laptop.}
    \label{fig: har} 
\end{figure}

\noindpar{Effect of audio.} With IMU only, mean accuracy over the 20 activities was 78.98\%. Adding the unmasked one second audio window raised it to 96.89\% with a standard deviation of 2.28 across folds. The gain of 17.91 points is large because of how the activity set was built. Figure \ref{fig: har} (A) shows the IMU-only confusion matrix. Errors lie almost entirely within groups: the four hammering surfaces are confused with each other, as are the four rackets and the four utensils. Across groups the IMU alone is nearly sufficient. In other words, the IMU does the coarse work and the audio does the fine-grained work, which is the division of labor the design intends. On a mixed set of everyday activities where most distinctions are coarse, the gain from audio would be smaller.

\begin{table}[t]
\centering
\begin{tabular}[t]{cccc}
\toprule
\textbf{Mask}&\textbf{Factor}&\textbf{Accuracy} &\textbf{Std.} \\
\hline
No Mask & - & 96.89\% & 2.28\%\\
\hline
\multirow{5}{*}{Point-wise Mask} & 50\% & 95.55\% & 2.72\%\\
                                 & 80\% & 95.17\% & 2.56\%\\
                                 & 90\% & 94.27\% & 3.15\%\\
                                 & 95\% & 92.23\% & 4.02\%\\
                                 & 97\% & 87.97\% & 4.14\%\\
\hline
\multirow{3}{*}{Contiguous Mask} & 50\% & 92.33\% & 3.82\%\\
                                 & 80\% & 88.19\% & 4.24\%\\
                                 & 90\% & 86.32\% & 3.92\%\\
\bottomrule
\end{tabular}
\caption{Recognition accuracy with IMU plus audio under different masking methods and factors. Mean per-activity accuracy over 20 activities and standard deviation across held-out participants.}
\label{tab: mask har}
\end{table}

\noindpar{Effect of masking.} Table \ref{tab: mask har} and Figure \ref{fig: results} (B) give accuracy under masking. Point-wise masking cost little. With 95\% of samples removed, accuracy was 92.23\%, and with 97\% removed it was 87.97\%. Compare this with Table \ref{tab: mask asr}, where the same masks raised speech recognition WER only to 40.33\% and 48.50\%. Activity sounds survive random sample dropout about as well as speech does, so point-wise masking does not buy privacy at the expense of activity information. It does not buy much privacy either.

Contiguous masking costs more. At 50\% accuracy fell to 92.33\%, at 80\% to 88.19\%, and at 90\% to 86.32\%. The drop is expected since a contiguous mask can remove the whole sound event from the window if the random start position falls on it. At the same time, 50\% contiguous masking pushed speech WER to 69.41\% in Table \ref{tab: mask asr}. The asymmetry is the useful finding. Cutting the microphone-on time to 100 ms per activation still left accuracy 7.34 points above IMU only, while the same cut was the most damaging to speech recognition of anything we tried.

\noindpar{Per-activity view.} Figure \ref{fig: har} shows confusion matrices for IMU only (A), audio only without masking (B), and IMU plus audio with a 10\% point-wise mask (C) and a 50\% contiguous mask (D). The within-group confusions of panel A largely disappear when audio is added, including under masking. The remaining errors under contiguous masking are concentrated in the sports group and in the utensil group, which are the groups where the trigger timing is least precise and the sound is most likely to fall in the masked portion of the window.

\section{Discussion} \label{sec: discussion}

\subsection{Effect of a Shorter Window} \label{sec: window}
Reviewers of an earlier version asked how performance depends on the window length. We did not run a full ablation, and the one comparison we have comes from an earlier training run whose one second baselines differ by up to a point from the final numbers in Section \ref{sec: HAR results}, so we report it as differences rather than absolute values. Halving both windows from one second to 0.5 seconds, with the same leave-one-participant-out protocol, lowered mean accuracy by about 3.7 points with IMU only, about 7.2 points with audio only, and about 5.2 points with IMU plus unmasked audio. Under masking the drop was between 4 and 6 points across the settings we ran. The audio branch lost more than the IMU branch, which is consistent with the timing picture in Section \ref{sec: chronology}: a shorter audio window starting at the anomaly point more often ends before the sound arrives. A shorter window is a smaller microphone exposure, so this trade-off is the one a deployment would tune, and a proper sweep over window length is the first experiment we would add.

\subsection{Non-Dominant Hand} 
Many two-handed actions involve the non-dominant hand in a supporting role, and a watch is often worn on that hand. As an informal check we asked one participant from the main study to perform serves in badminton, tennis, ping-pong, and pickleball with the watch on the non-dominant wrist, five serves per sport in each of two sessions with the watch removed between sessions, for 10 serves per sport. We trained on eight serves per sport and tested on the remaining two, with a 90\% contiguous mask, and obtained 97.22\% accuracy on the segments from the held-out serves. With one participant and a within-participant split this is an anecdote, not a result, and we include it only to indicate that the trigger fires on the supporting hand as well.

\subsection{Unseen Activities} \label{sec: unseen}
\begin{figure}[b]
    \centering
    \makebox[0pt]{\includegraphics[width=\columnwidth]{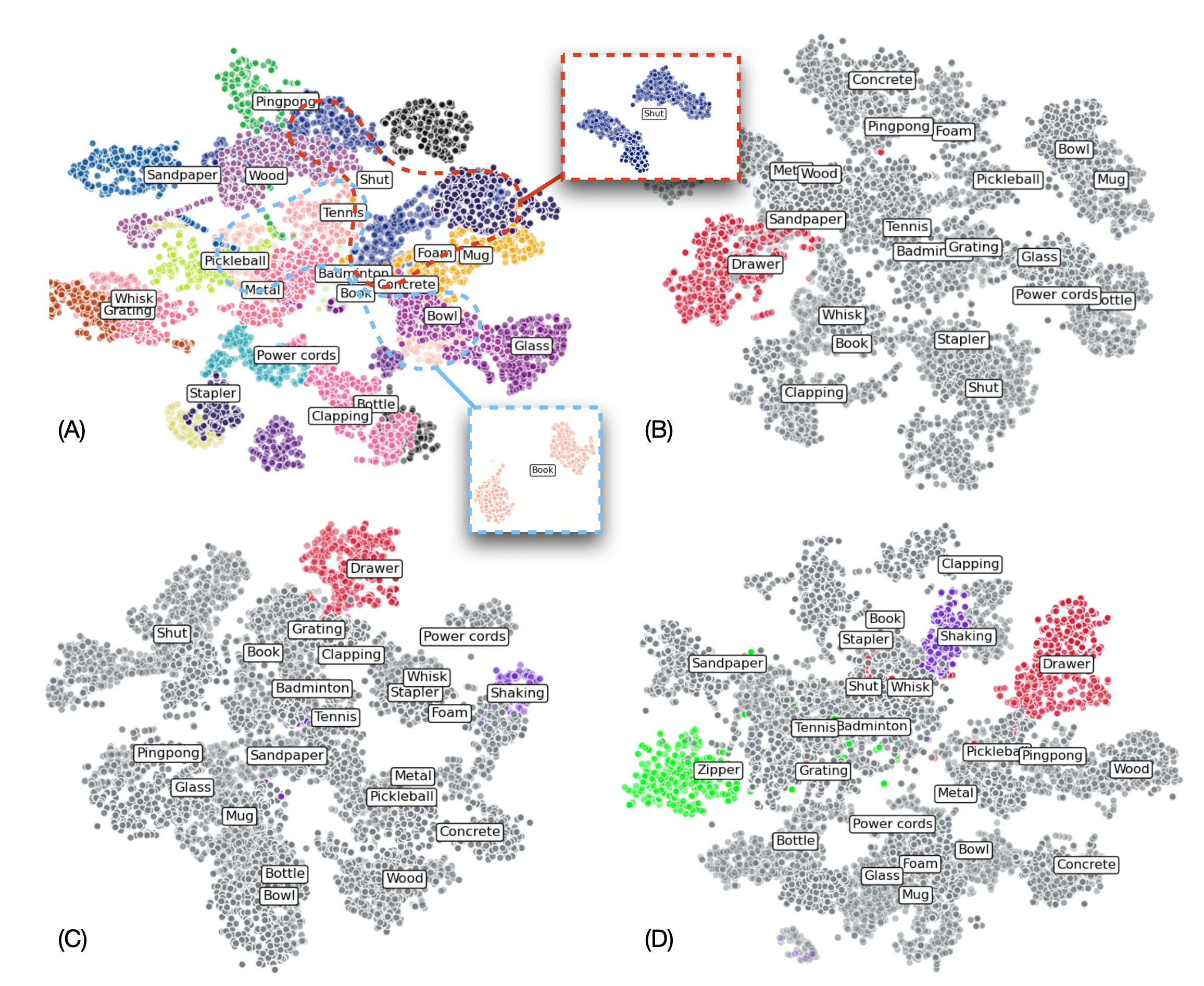}}
    \caption{t-SNE plots of KMeans clusters over fused features. (A) The 20 studied activities. (B) After adding the drawer activity. (C) After also adding the correction fluid activity. (D) After also adding the zipper activity. Each new activity appears as a separate cluster.}
    \Description{Four two-dimensional scatter plots with points colored by cluster and each cluster annotated with an activity name. Plot A shows about twenty colored clusters. Plots B, C, and D each add one further cluster, labeled drawer, correction fluid, and zipper, that sits apart from the existing clusters rather than overlapping them.}
    \label{fig: cluster}
\end{figure}

The recognizer in the main study is supervised and needs labeled data per activity. The trigger is not, and one consequence is that it fires on activities the recognizer has never seen. To look at what happens to those segments we recorded three additional activities from one participant: opening and closing a drawer, shaking a bottle of correction fluid, and pulling a zipper on a bag. Each was performed 10 times per session over 10 sessions. We ran KMeans on the recognizer's fused features and plotted the result with t-SNE. Figure \ref{fig: cluster} (A) shows the 20 studied activities, and (B) through (D) show each new activity added in turn. The new activities form clusters of their own rather than merging into existing ones. This suggests a workflow in which the trigger and the feature extractor run on unseen activities, segments accumulate in new clusters, and a user labels each cluster once rather than each segment. Unsupervised IMU HAR methods such as KMeans and VAEs have been used this way before \cite{chang2020systematic, moschetti2017daily, bai2019motion2vector}. We have not built or evaluated the labeling workflow, and the plot is one participant.

\subsection{Toward a Live Deployment}
Three things separate our offline evaluation from a system running on a watch.

\noindpar{Latency.} The detector evaluates a one second window every 0.1 seconds, so the earliest an anomaly can be reported is about 0.1 seconds after the motion that causes it, plus inference time. For impulsive sounds that arrive at the moment of the velocity change, part of the onset can be lost. Two mitigations are possible without keeping the microphone on. One is to lower the detector threshold so it fires earlier in the rising edge of the error, at the cost of precision. The other is to keep a short audio ring buffer of a few hundred milliseconds in device memory that is discarded unless a trigger arrives. The second option weakens the argument that no audio exists before the trigger, and a deployment would have to decide whether a buffer that never leaves the sensing layer is acceptable under its threat model. We did not implement either.

\noindpar{Ambient noise.} Our sessions were quiet. In daily life the one second window will often contain conversation, traffic, or appliance noise alongside the activity sound. This affects both the privacy side and the recognition side. On the privacy side, a window that contains speech is a window that leaks speech, and the trigger has no way to know this. On the recognition side, the audio branch was trained on clean sound and will degrade. The standard mitigations, training with noise augmentation and rejecting windows whose energy profile does not match a short impulsive event, are applicable to the audio branch and we have not tested them. Noise is the largest unaddressed factor between these results and a working system.

\noindpar{Unscripted motion.} The recorded sessions contain little unscripted motion, so our false trigger rate is measured against a narrow background. A day of ordinary wrist motion will produce more anomalies than a 30 minute scripted session, and each false trigger is a one second microphone activation. The detector's absolute error floor and its threshold are the two knobs that control this, and both would need to be set against a day-long recording of ordinary motion, which we did not collect.

\subsection{Improving the Detector}
The detector is a single-layer LSTM-VAE, and its recall of 74.28\% leaves room. Transformer-based anomaly detectors such as the Anomaly Transformer \cite{xu2021anomaly} model longer context than a single-layer LSTM and could help with the sports group, where the anomaly is spread over a long swing. The detector could also be trained on a broader set of ordinary motion than our 10 activities. We chose the LSTM-VAE for its small size, since the detector must run continuously on the watch, and any replacement has to fit the same budget.

\section{Limitations and Future Work} \label{sec: limitations}

\noindpar{The activity set favors audio.} The 20 activities were chosen so that within-group distinctions require sound. This is the case the design targets, but it also means the reported gain from audio over IMU is an upper end. On a set that mixes coarse and fine-grained distinctions, IMU-only accuracy would be higher and the gap smaller. Some groups, the hammering surfaces in particular, are stress tests rather than tasks a user would ask a watch to recognize. We see the results as evidence that short, masked audio windows carry enough information to resolve fine-grained distinctions, and not as an estimate of accuracy on daily life.

\noindpar{Why we did not use an existing dataset.} We would have preferred to validate on public data. The wearable datasets with both IMU and audio that we know of either release audio at a low sampling rate, as SAMoSA does at 1 kHz, or contain coarse activities that the IMU already separates. Neither lets us test the fine-grained case with full-rate audio. Releasing our dataset is our attempt to make the next comparison easier.

\noindpar{Ground truth for the trigger comes from the audio.} Sound events were defined by an energy threshold on the recorded audio, not by video or manual annotation. Quiet events can be missed by the threshold and loud non-activity sounds can be counted as events, and we did not verify the event timeline by hand. The precision and recall in Section \ref{sec: segmentation results} should be read with this in mind.

\noindpar{Controlled sessions and sample size.} Fifteen participants in 30 minute scripted sessions is enough to show feasibility and not enough to characterize variation across people, environments, or time. Sessions contained little unscripted motion and no ambient noise. The masking check used four speakers and a phone rather than the watch. The non-dominant hand and unseen activity observations are from one participant. All of these are stated where they appear and none should be generalized.

\noindpar{Inclusiveness of the activities.} The sports group assumes some familiarity with racket sports, and one participant could not complete the badminton activity. The set as a whole reflects the authors' surroundings and is not balanced across age, gender, or occupation. An activity set built from a survey of what users actually do would be more representative.

\noindpar{Compound and surface-dependent sounds.} Some actions produce sounds from more than one source. Cutting a vegetable produces a sound from the vegetable and a louder one from the knife meeting the board, and the second can mask the first. The same action on different surfaces, placing a bowl on wood, glass, or metal, produces different sounds, and our recognizer, trained on one table, would confuse the surface with the action. Both cases would need training data that varies the surface, or a model that separates action from surface.

\noindpar{The audio window is not aligned to the sound.} The system requests one second of audio at the anomaly point. It does not predict when inside that second the sound will occur or how long it will last. A detector that also predicted the sound onset would allow shorter windows and less exposure. Related to this, some activities produce sound only on part of the action. Opening a laptop is silent and closing it is not, and the clustering in Figure \ref{fig: cluster} (A) splits the laptop activity into two clusters accordingly. Treating open and close as separate labels would be more honest than our single label.

\noindpar{Privacy is reduced, not guaranteed.} Section \ref{sec: threat model} states this and we repeat it here. Short, masked, IMU-triggered windows reduce how much speech is recorded. They do not prevent an adversary who accumulates many windows, or who applies audio inpainting, from recovering some content. A formal intelligibility study of the kind SAMoSA ran, with human listeners and multiple recognizers, is needed before any stronger claim, and IMU privacy is not addressed at all.

\section{Conclusion}
We presented \systemname{}, a microphone activation approach for wrist wearables in which an unsupervised IMU anomaly detector decides when to turn the microphone on, for at most one second, and the captured second is masked before recognition. On a controlled dataset of 20 fine-grained activities from 15 participants, the trigger reached 86.46\% precision and 74.28\% recall against sound events, and recognition with IMU plus masked audio stayed at 86.32\% with the microphone effectively on for 100 ms per activation, against 78.98\% for IMU alone. The evaluation is offline and the activity set was built to favor audio, so the numbers are evidence of feasibility rather than of field performance. The threat model, the measured trigger accuracy, and the asymmetry between what contiguous masking does to speech recognition and what it does to activity recognition are the parts we expect to carry over to a deployed system. Noise, unscripted motion, and a formal intelligibility study are the parts still missing.

\bibliographystyle{ACM-Reference-Format}
\bibliography{ref.bib}

\end{document}